\documentclass[sigconf,natbib=true]{acmart}

\AtBeginDocument{
  }

\setcopyright{acmlicensed}
\copyrightyear{2026}
\acmYear{2026}

\acmConference[Conference acronym 'XX]{Make sure to enter the correct
  conference title from your rights confirmation email}{June 03--05,
  2018}{Woodstock, NY}

\acmISBN{978-1-4503-XXXX-X/2026/07}

\usepackage[utf8]{inputenc}
\usepackage{url}

\usepackage{amssymb}
\usepackage{booktabs}
\usepackage{bbding}
\usepackage{pifont}
\usepackage{wasysym}
\usepackage{utfsym}
\usepackage{fontawesome}
\usepackage{multirow}

\begin{document}

\title{AesRec: A Dataset for Aesthetics-Aligned Clothing Outfit Recommendation}

\author{Wenxin Ye}
\email{wenxinye@whut.edu.cn}
\affiliation{
  \institution{Wuhan University of Technology}
  \country{China}
}

\author{Lin Li}
\email{cathylilin@whut.edu.cn}
\affiliation{
  \institution{Wuhan University of Technology}
  \country{China}
}

\author{Ming Li}
\email{mingli7@yorku.ca}
\affiliation{
  \institution{York University}
  \country{Canada}
}
\affiliation{
  \institution{Wuhan University of Technology}
  \country{China}
}

\author{Yang Shen}
\email{353939@whut.edu.cn}
\affiliation{
  \institution{Wuhan University of Technology}
  \country{China}
}

\author{Kanghong Wang}
\email{wangkanghong03@whut.edu.cn}
\affiliation{
  \institution{Wuhan University of Technology}
  \country{China}
}

\author{Jimmy Xiangji Huang}
\email{jhuang@yorku.ca}
\affiliation{
  \institution{York University}
  \country{Canada}
}

\renewcommand{\shortauthors}{Ye et al.}

\begin{abstract}

Clothing recommendation extends beyond merely generating personalized outfits; it serves as a crucial medium for aesthetic guidance. However, existing methods predominantly rely on user-item-outfit interaction behaviors while overlooking explicit representations of clothing aesthetics. To bridge this gap, we present the AesRec benchmark dataset featuring systematic quantitative aesthetic annotations, thereby enabling the development of aesthetics-aligned recommendation systems. 
Grounded in professional apparel quality standards and fashion aesthetic principles, we define a multidimensional set of indicators. At the item level, six dimensions are independently assessed: silhouette, chromaticity, materiality, craftsmanship, wearability, and item-level impression. Transitioning to the outfit level, the evaluation retains the first five core attributes while introducing stylistic synergy, visual harmony, and outfit-level impression as distinct metrics to capture the collective aesthetic impact.
Given the increasing human-like proficiency of Vision-Language Models in multimodal understanding and interaction, we leverage them for large-scale aesthetic scoring. We conduct rigorous human-machine consistency validation on a fashion dataset, confirming the reliability of the generated ratings. Experimental results based on AesRec further demonstrate that integrating quantified aesthetic information into clothing recommendation models can provide aesthetic guidance for users while fulfilling their personalized requirements.

\end{abstract}




\maketitle

\section{Introduction}

In recent years, intelligent recommendation technologies are evolving beyond mere preference matching to actively guide users toward higher-quality choices, particularly in aesthetic domains such as fashion~\cite{zhang2018deep,amershi2019guidelines}. This evolution aligns with the "AI for Good"~\cite{holstein2019improving,floridi2018ai4people,floridi2021design} philosophy, leveraging AI to provide aesthetic guidance in tandem with personalized recommendations.

\begin{figure}[htbp]
  \centering
  \includegraphics[width=0.45\textwidth]{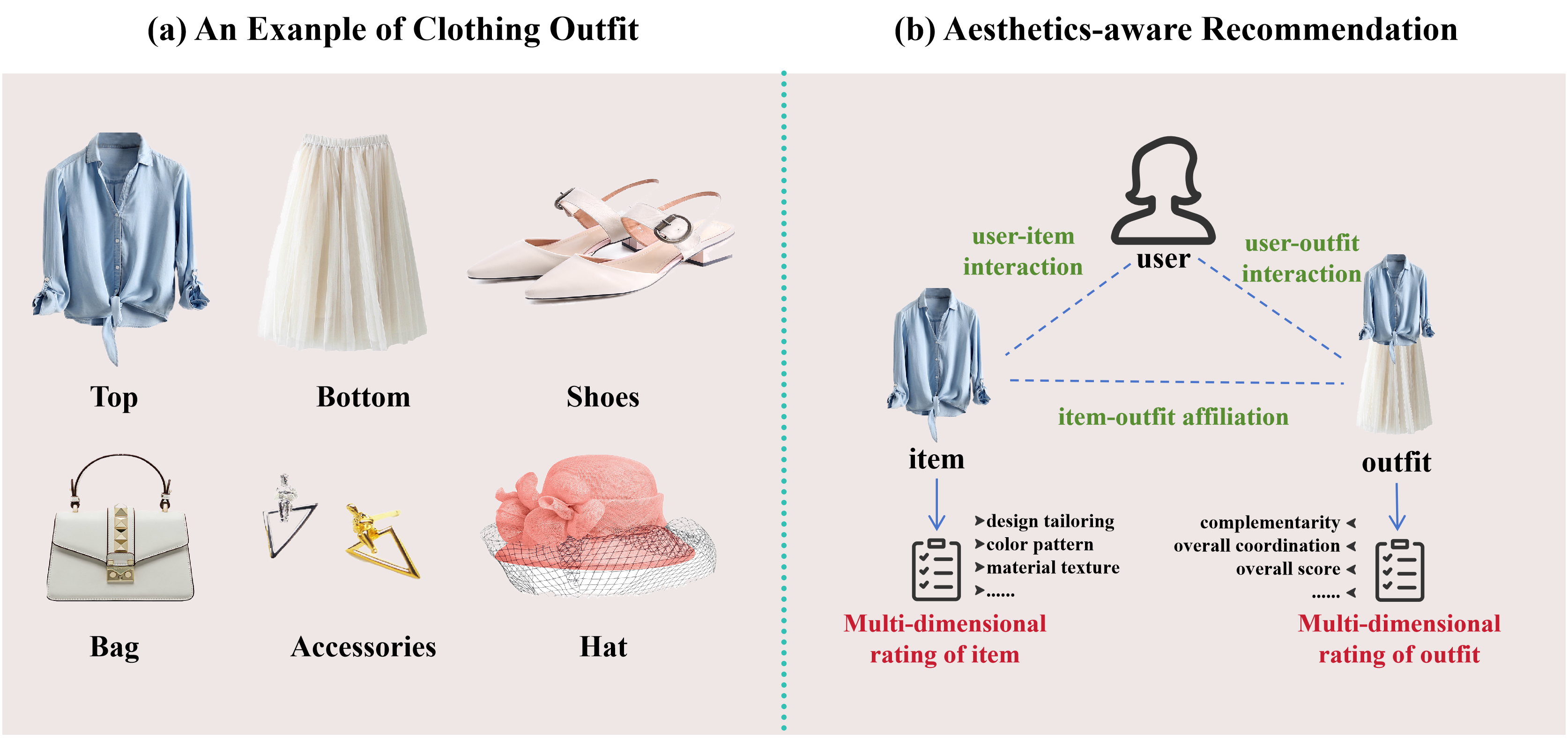}
  \caption{(a) An Example of Clothing Outfit; (b) Aesthetics-oriented Recommendation Relationship}
  \label{fig:outfit_example}
  \end{figure}

  \begin{table*}[t]
    \centering
    \caption{Comparison of existing clothing recommendation datasets}
    \label{tab:dataset_comparison}
    \resizebox{\textwidth}{!}{
    \begin{tabular}{@{}llclp{2.5cm}cccccc@{}}
    \toprule
    Dataset & Reference & Year & \multicolumn{1}{c}{Data Source} & Outfit-Item & User-Outfit & User-Item & Category & Aesthetic & Open \\
     & & & & Affiliation & Interaction & Interaction & Labels & Annotation & Source \\
    \midrule
    DeepFashion & \cite{liu2016deepfashion} & 2016 & E-commerce \& Consumer Photos & \texttimes & \texttimes & \texttimes & \checkmark (50) & \texttimes & \checkmark \\
    \midrule
    Fashionpedia & \cite{jia2020fashionpedia} & 2020 & Daily, Street, Celebrity, Fashion Shows & \texttimes & \texttimes & \texttimes & \checkmark (27) & \texttimes & \checkmark \\
    \midrule
    Vibrent Rental & \cite{borgersen2024vibrent} & 2024 & Fashion Rental Platform & \texttimes & \texttimes & \checkmark (77.1k) & \checkmark & \texttimes & \checkmark \\
    \midrule
    POG & \cite{chen2019pog} & 2019 & Alibaba iFashion & \checkmark (1.01M) & \checkmark & \checkmark & \checkmark (80) & \texttimes & \checkmark \\
    \midrule
    FLORA & \cite{deshmukh2024flora} & 2024 & Generative Model Auto-construction & \checkmark (4,330 pairs) & \texttimes & \texttimes & \texttimes & \texttimes & \checkmark \\
    \midrule
    Aesthetics Fashion & \cite{gaur2014ranking} & 2014 & Surrey University & \texttimes & \texttimes & \texttimes & \checkmark (11+4) & \checkmark (Pairwise) & \checkmark \\
    \bottomrule
    \end{tabular}
    }
    \end{table*}

  Aesthetic-aligned recommendations are an important direction in the field of clothing recommendation, which explicitly quantifies and integrates aesthetic quality into the recommendation process. In traditional clothing recommendation approaches, as illustrated in Figure~\ref{fig:outfit_example}a, systems primarily optimize for user preference matching based on historical interactions, where a complete clothing outfit typically comprises items from multiple categories but lacks explicit aesthetic guidance. In contrast, aesthetics-aligned systems elevate user experiences by promoting aesthetically superior content while maintaining personalized relevance. As demonstrated in Figure~\ref{fig:outfit_example}b, the data relationships in aesthetics-aligned clothing outfit recommendation integrate quantitative aesthetic annotations into the recommendation framework, enabling the system to guide users toward higher-quality outfit choices that balance both personal preferences and aesthetic excellence.

However, \textbf{existing clothing recommendation datasets universally do not include indicators of clothing aesthetic quality}, which constrains the advancement of aesthetics-aligned recommendation systems. On the one hand, interaction-based methods mainly emphasize personalization by learning user preferences from historical interactions~\cite{rendle2009bpr,he2020bgcn,chen2020lightgcn,wu2021self}, but they fail to achieve aesthetic alignment without explicit aesthetic annotations. On the other hand, vision-based methods focus on visual compatibility modeling~\cite{vasileva2018learning,han2017learning,hsiao2021outfit}, but often fail to provide personalized recommendations due to the absence of interactive data. Therefore, existing work has not adequately addressed the issue of aesthetic and personalization compatibility.

 It remains a challenging task to collect large-scale quantitative aesthetic data, largely due to the prohibitive costs associated with traditional manual labeling. This difficulty is compounded by the need to reconcile individual subjective variances with universal objective standards within a multidimensional annotation.In Table \ref{tab:dataset_comparison} of Section 2.1, we systematically compared six existing datasets. In terms of scale, the Aesthetics Fashion dataset, the only one containing aesthetic annotations, contains only 1,064 images, a difference of 2-3 orders of magnitude compared to other mainstream datasets (DeepFashion with 800,000+ images, POG with 1.01 million outfits). The root of this bottleneck lies in the labor-intensive nature of traditional manual annotation methods: constructing a million-level aesthetic dataset requires thousands of person-years of work, which is difficult to achieve in practice. As shown in Figure \ref{fig:outfit_example}b, both individual items and sets require scoring across multiple dimensions, including color harmony and style consistency. Each dimension requires a balance between subjective aesthetic preferences and objective design standards, placing extremely high demands on the annotators' professional abilities and consistency across dimensions. This makes the acquisition of high-quality, large-scale aesthetic data a key bottleneck in aesthetic-aligned recommender system research.

To facilitate clothing matching recommendations that align with aesthetic principles, we construct the \textbf{AesRec benchmark dataset} with quantitative aesthetic indicators. The construction of this dataset comprises two key parts: the design of multidimensional aesthetic indicators and large-scale aesthetic assessment. \textit{1)} We design multidimensional aesthetic indicators following internationally recognized clothing quality standards and established aesthetic principles in fashion and style research. Specifically, for item level, six dimensions are independently assessed: silhouette, chromaticity, materiality, craftsmanship, wearability, and item-level impression. At the outfit level, the evaluation retains the first five core attributes while introducing stylistic synergy, visual harmony, and outfit-level impression as distinct metrics to capture the collective aesthetic impact. \textit{2)} Given the increasing human-like capabilities of Vision-Language Models (VLMs), we use a VLM with hundreds of millions of parameters for large-scale aesthetic scoring. Furthermore, we extract 100 sets of outfits and have three experts in the field of clothing manually score them. The average of the expert scores is used, and indicators such as the Pearson correlation coefficient are employed to verify the consistency between the manual scoring and the large-scale model's scores.

Based on this benchmark, we compare seven mainstream bundle recommendation models. Evaluation across two metrics, including average aesthetic score and exposure rank, demonstrates that integrating aesthetic guidance through loss functions significantly enhances the aesthetic quality of recommendations. These findings provide empirical evidence for the design of aesthetics-aware clothing recommendation systems.

\textbf{The main contributions of this paper include}: 1)constructing the AesRec dataset, filling the critical gap of lacking aesthetic indicators in the clothing recommendation domain; 2) validating the effectiveness of incorporating aesthetic alignment in recommendation models, demonstrating that aesthetic indicators can significantly improve the aesthetic quality of the recommendation results.

\section{Related Work}

\subsection{Datasets for Outfit Recommendation}

We systematically compare part of the existing datasets in Table~\ref{tab:dataset_comparison}, which can be divided into two categories: those with aesthetic annotations and those without aesthetic annotations.

\textbf{Datasets without Aesthetic Annotations:} DeepFashion~\cite{liu2016deepfashion} contains over 800k images with 50 categories and 1,000 attributes, focusing on recognition and retrieval tasks. Fashionpedia~\cite{jia2020fashionpedia} provides 48,825 images with instance segmentation and 294 fine-grained attributes across 27 apparel categories. POG~\cite{chen2019pog} scales to 1.01 million outfits and 280 million user interactions for personalized recommendation, while Vibrent Rental~\cite{borgersen2024vibrent} introduces 77.1k transactions in rental scenarios with temporal dynamics. FLORA~\cite{deshmukh2024flora} offers 4,330 outfit-text pairs for generative modeling with professional fashion terminology. These datasets focus on personalized outfits recommendations and do not include aesthetic labels.

\textbf{Datasets with Aesthetic Annotations:} The Aesthetics Fashion dataset~\cite{gaur2014ranking} provides systematic aesthetic evaluation, containing 1,064 images ranked through pairwise comparisons by fashion experts and crowdsourced annotators across 11 clothing and 4 body shape categories. The annotation process required 57,400 pairwise judgments and employed the Kemeny-Young aggregation method to generate global rankings. Although this method includes aesthetic annotations, it lacks personalized interaction data.

Currently, some large-scale datasets offer personalized interactions but lack aesthetic annotations, while datasets with aesthetic annotations lack personalized user interaction data. This hinders the development of clothing recommendation systems that are both aesthetically aligned and personalized.

\subsection{Methods for Outfit Recommendation}

Existing methodologies can be categorized into interaction-based and vision-based approaches

\textbf{Interaction-based Methods:} These treat outfits as fundamental recommendation units. Early models like BPR-MF\cite{rendle2009bpr} and BGCN\cite{he2020bgcn} focus on independent modeling of hierarchical preferences. Recent frameworks, including MultiCBR\cite{ma2024multicbr}, HyperMBR\cite{ke2023hyperbolic}, CrossCBR\cite{ma2022crosscbr}, and MIDGN\cite{zhao2022midgn}, leverage contrastive learning and graph architectures to facilitate collaborative information sharing between item and outfit levels~\cite{chen2020lightgcn,wu2021self}. However, these methods essentially optimize for user preference fitting rather than aesthetic excellence.

\textbf{Vision-based Methods:} These focus on visual compatibility modeling. Beyond early Siamese CNN architectures for pairwise matching~\cite{vasileva2018learning,han2017learning}, modern approaches employ GNNs or Large Language Models (e.g., LLaVA, Qwen) to integrate semantic fashion knowledge into outfit generation~\cite{radford2021learning,liu2023llava,wei2023zeroshot,hou2024large}. Despite their visual focus, they often conflate "compatibility" with "aesthetic quality" and rely on implicit heuristic criteria, making systematic aesthetic improvement difficult.

Despite their effectiveness, existing methods generally fail to provide a structured and multidimensional formulation of aesthetic quality. Interaction-based approaches implicitly treat user preference as a proxy for aesthetics, which may reinforce subjective or historically biased taste patterns. Vision-based generative methods, while modeling visual compatibility, often conflate compatibility with aesthetic quality and rely on heuristic or implicit criteria, making it difficult to systematically evaluate or improve the aesthetic level of generated outfits.

\section{Dataset Construction}

The development of our dataset involves several key steps, including data source selection, preprocessing, indicator design, and the use of VLM for aesthetic assessment. Consequently, this section provides a detailed exposition of these core components.

\begin{table}[htbp]
\centering
\caption{Description of the POG dataset}
\label{tab:pog_structure}
\scalebox{0.8}{
\begin{tabular}{@{}lll@{}}
\toprule
\multicolumn{3}{c}{\textbf{Outfit data in the POG dataset}} \\
\midrule
Field & Description & Example \\
\midrule
outfit\_id & outfit identifier & 0x12345 \\
item\_id & item identifier & 0x67890 \\
item\_ids & list of item IDs in the outfit & 0x67890;0x67891;0x67892 \\
category\_id & Category ID of the item & 0x154567 \\
image\_url & URL of the item image & http://img.alicdn.com/... \\
title & description of the item & Women's Casual T-shirt \\
\midrule
\multicolumn{3}{c}{\textbf{User interaction in the POG dataset}} \\
\midrule
Field & Description & Example \\
\midrule
user\_id & user identifier & 0x345678 \\
outfit\_id & outfit identifier & 0x123455 \\
interaction\_type & type of interaction (click/view) & click \\
\bottomrule
\end{tabular}}
\end{table} 

\subsection{Data Sources and Preprocessing}

\textbf{Data Sources:} Based on a comparative analysis of existing datasets in Table~\ref{tab:dataset_comparison}, we find that the POG dataset provides diverse and extensive data support—encompassing multimodal information across images, text, and interactions, such as explicit item-outfit correlations, rich item-level metadata, and large-scale user logs—which perfectly aligns with the requirements for constructing a multidimensional aesthetic benchmark. Consequently, this study selects the POG dataset as the primary data source. This dataset originates from iFashion, the fashion recommendation platform of Alibaba's Taobao, and has significant practical research value. Its structural details are summarized in Table~\ref{tab:pog_structure}.

The POG dataset consists of three components: (1) \textbf{Outfit data} containing 1.01 million manually curated outfits composed of 583K items, where each outfit has been reviewed by fashion experts before publication; (2) \textbf{User interaction data} capturing 280 million click behaviors from 3.57 million active users, recording both user-outfit and user-item interactions over a three-month period; (3) \textbf{Item pool} with 4.75 million items covering 80 fine-grained fashion categories. Each item is annotated with white-background product images, textual descriptions, and category labels, enabling both visual and semantic modeling. This holistic architecture—simultaneously capturing user preferences at both outfit and item levels—makes POG particularly suitable for our aesthetic-aware recommendation research.

\textbf{Data preprocessing:} We validate all image URLs and remove items with invalid image resources, along with the outfits containing them, to ensure that all remaining items have valid visual inputs for aesthetic scoring. In total, 37 items with invalid image URLs are removed. After data cleaning, we convert all identifiers into continuous integer IDs and construct a unified user-outfit-item graph. The original POG dataset categorizes items into 80 fine-grained categories without semantic labels. To improve interpretability and consistency with clothing outfit analysis, we manually consolidate these categories into 17 major categories, covering the main components of outfit composition such as tops, bottoms, bags, and accessories.

\subsection{Multi-dimensional Aesthetic Indicators}

Since there is currently no unified standard for judging clothing aesthetics, we construct a multidimensional aesthetic index as shown in Table \ref{indicators}, based on internationally recognized clothing quality standards~\cite{kawabata1980standardization,kawabata1975objective,cooklin2006introduction} and established aesthetic principles in fashion style research~\cite{aldrich2015metric,joseph2014patternmaking,itten1973art,fan2014clothing,hunter2004clothing,vasileva2018learning,han2017learning,murray2012ava}, combined with the suggestions of experts in the field of clothing.

\begin{table}[h]
\centering
\caption{Multi-dimensional aesthetic indicators and scoring criteria.}
\label{indicators}
\small
\resizebox{0.95\columnwidth}{!}{
\begin{tabular}{p{0.25\columnwidth} p{0.70\columnwidth}}
\toprule
\textbf{Indicator} & \textbf{Description} \\
\midrule
Silhouette & Evaluates silhouette structure, proportional balance, and cutting quality. \\
Chromaticity & Assesses color harmony, pattern organization, and overall visual balance. \\
Materiality & Reflects fabric quality, texture clarity, and material expressiveness. \\
Craftsmanship & Focuses on construction quality, stitching, finishing, and detail execution. \\
Wearability & Captures wearability, comfort, and suitability for intended scenarios. \\
Item-level impression(II) & Provides a holistic item-level aesthetic judgment. \\
\midrule
Stylistic Synergy & Evaluates inter-item compatibility in style, color, material, and design. \\
Visual Harmony & Assesses global unity, coherence, and visual harmony of the outfit. \\
Outfit-level impression(OI) & Provides a holistic outfit-level aesthetic judgment. \\
\bottomrule
\end{tabular}
}
\end{table}

At the item level, we define six evaluation dimensions, which can be categorized into three groups: (1) \textbf{Visual aesthetic dimensions} (Silhouette, Chromaticity, Materiality) capture visual appearance and aesthetic effects arising from structural design, color harmony, and fabric characteristics. These dimensions draw on classical garment construction principles~\cite{aldrich2015metric,joseph2014patternmaking}, color theory foundations~\cite{itten1973art}, and textile aesthetics research~\cite{kawabata1980standardization,kawabata1975objective} that establish visual elements as fundamental determinants of apparel aesthetics. (2) \textbf{Quality assessment dimensions} (Craftsmanship, Wearability) evaluate construction quality, finishing precision, and functional aspects such as comfort and wearability, following apparel quality evaluation practices~\cite{cooklin2006introduction} and international quality standards\footnote{International Organization for Standardization (ISO), American Society for Testing and Materials (ASTM), and American Association of Textile Chemists and Colorists (AATCC) establish standards for apparel quality evaluation. See \url{https://www.iso.org}, \url{https://www.astm.org}, and \url{https://www.aatcc.org} for details.}, as well as ergonomics research~\cite{fan2014clothing,hunter2004clothing} that emphasize workmanship and functional suitability as core indicators of clothing quality. (3) \textbf{Item-level Overall evaluation} (Item-level impression) provides a holistic aesthetic judgment at the item level, following aesthetic evaluation paradigms adopted in large-scale visual aesthetic datasets~\cite{murray2012ava}, where global aesthetic ratings complement fine-grained attribute annotations.

At the outfit level, combining the five indicators of visual aesthetic dimensions and quality assessment dimensions, we added three more dimensions: \textbf{Stylistic Synergy} evaluates inter-item compatibility and aesthetic harmony, informed by fashion compatibility modeling studies~\cite{vasileva2018learning,han2017learning} that explicitly treat inter-item compatibility as a distinct aesthetic factor. \textbf{Visual Harmony} assesses global visual unity and coherence, consistent with holistic outfit assessment practices in fashion styling theory and design evaluation. \textbf{Outfit-level impression} provides a holistic aesthetic evaluation of a garment set. It represents a synthesized manifestation of various aesthetic attributes, reflecting the overall aesthetic.

\subsection{Human-Machine Consistency Validation}

Annotating large-scale datasets across multiple dimensions is highly resource-intensive in terms of both time and expense. Meanwhile, evaluations conducted exclusively by LLMs often struggle with reliability issues. Therefore, we conduct a rigorous scoring consistency assessment to verify the consistency between VLM scoring and human annotation.

\subsubsection{Validation Design} We adopt a two-stage progressive validation design~\cite{schwarz2018will,deng2017image}. First, we conduct pre-validation on the Photo Critique Captioning Dataset (PCCD), an established photography aesthetics benchmark containing images with multidimensional manual aesthetic scores, to evaluate the basic capability of LLMs in aesthetic assessment. Second, we perform domain-specific validation on fashion samples to confirm model effectiveness for clothing aesthetic assessment. This hierarchical design ensures both generalizability and domain applicability. 

\subsubsection{Data Debiasing}

Due to potential systematic discrepancies in scoring scales between VLMs and human annotators—where VLMs may exhibit an overall bias toward higher or lower ratings—comparing raw scores directly may fail to accurately capture the underlying correlation. To address this, we employ a standardized de-biasing method: for each evaluation dimension, the mean ($\mu_{model}$, $\mu_{human}$) and standard deviation ($\sigma_{model}$, $\sigma_{human}$) of both model and human scores are calculated. We then apply a linear transformation to the model scores, as shown in Equation \ref{eq1}, to align their distribution with that of the human ratings. This normalization process eliminates systematic bias, enabling a more precise assessment of consistency in terms of ranking and relative relationships.

\begin{equation}
 score_{normalized} = \frac{score_{model} - \mu_{model}}{\sigma_{model}} \times \sigma_{human} + \mu_{human}
 \label{eq1}
\end{equation}

\subsubsection{Validation Metrics} To comprehensively assess human-machine agreement, we employ multiple consistency metrics. \textit{Correlation metrics} include: Pearson correlation coefficient ($r$) to measure linear relationship strength and Spearman rank correlation ($\rho$) to measure ranking consistency with robustness to outliers. \textit{Error metrics} include: Mean Absolute Error (MAE) to quantify average deviation magnitude and Root Mean Square Error (RMSE) to measure deviation with sensitivity to large discrepancies. These complementary metrics provide comprehensive inter-rater reliability assessment from different perspectives.

\subsubsection{Validation on the Photographic Aesthetics} 
We randomly sample 100 images from the PCCD and use Qwen-VL-Max\footnote{Accessed via aliyun at \url{https://bailian.console.aliyun.com}} to score the dataset based on predefined aesthetic dimensions. After calibrating the scores to align the mean and variance of the model scores and human annotations, we calculate the evaluation metrics. As shown in Table~\ref{tab:pccd_validation}, the model achieves strong agreement with human annotations across all metrics: Pearson $r=0.847$ and Spearman $\rho=0.823$. Error metrics show MAE=0.523 and RMSE=0.687, indicating small deviations. These results demonstrate that \textbf{VLMs possess reliable capability in aesthetic assessment tasks}, establishing a foundation for application in the fashion domain. 

\begin{table}[htbp]
  \centering
  \caption{Consistency validation results on PCCD}
  \label{tab:pccd_validation}
  \scalebox{0.8}{
    \begin{tabular}{@{}lcccc@{}}
      \toprule
       Dimensions & Pearson $r$ & Spearman $\rho$ & MAE & RMSE \\
      \midrule
      Overall Impression & 0.8343 & 0.8816  & 0.5344 & 0.6924 \\
      Composition and Perspective & 0.9098 & 0.9185  & 0.5160 & 0.6466 \\
      Color and Lighting & 0.8897 & 0.8765  & 0.5311 & 0.6356 \\
      Subject of Photo & 0.8599 & 0.8426  & 0.6150 & 0.6852 \\
      Depth of Field & 0.9309 & 0.9482  & 0.4084 & 0.5965 \\
      Focus & 0.8440 & 0.8029  & 0.6425 & 0.7853 \\
      Exposure and Speed & 0.9114 & 0.9522  & 0.5179 & 0.6544 \\
      \bottomrule
      \end{tabular}}
  \end{table}

\subsubsection{Domain-Specific Validation on Fashion Samples.} 
For the fashion-specific validation, we recruit three independent expert annotators with specialized knowledge in fashion aesthetics. Each annotator scores 100 randomly sampled outfits across all eight dimensions (Design Tailoring, Color Pattern, Material Texture, Craftsmanship Detail, Practical Comfort, Complementary Matching, Overall Coordination, and Overall Score) on a 1.0-10.0 scale with one decimal precision. Annotators complete their assessments independently without mutual consultation to ensure annotation independence and avoid anchoring bias. The final ground truth for each dimension is computed as the arithmetic mean of three expert ratings, following established practices in subjective quality assessment~\cite{murray2012ava}. This multi-rater averaging effectively reduces variance from individual annotator subjectivity while maintaining inter-rater reliability.

For these 100 sets, Qwen-VL-Max and Gemini-3-pro\footnote{Accessed via Google AI Studio at \url{https://aistudio.google.com/}} are used to score the item-level in 6 dimensions and the outfit-level in 8 dimensions, and evaluation indicators are calculated after the scores are calibrated. Table~\ref{tab:item_validation_comparison} and table \ref{tab:validation_comparison} present the inter-rater agreement results. Qwen-VL-Max demonstrates strong agreement with expert annotations: Pearson $r=0.812$, Spearman $\rho=0.798$, with MAE=0.634 and RMSE=0.823. Gemini shows comparable performance with Pearson $r=0.793$ and Spearman $\rho=0.781$, though with slightly higher errors (MAE=0.687, RMSE=0.891). Both models achieve correlation coefficients above 0.75 across all metrics, indicating reliable inter-rater reliability in the fashion domain. Qwen-VL-Max's marginally superior performance across both correlation and error metrics makes it the preferred choice for large-scale annotation.

\begin{table}[htbp]
  \centering
  \caption{Item-level consistency validation results }
  \label{tab:item_validation_comparison}
  \scalebox{0.7}{
  \begin{tabular}{lcccccccc}
  \toprule
  \multirow{2}{*}{Dimensions} & \multicolumn{2}{c}{Pearson $r$} & \multicolumn{2}{c}{Spearman $\rho$} & \multicolumn{2}{c}{MAE} & \multicolumn{2}{c}{RMSE} \\
  \cmidrule(lr){2-3}\cmidrule(lr){4-5}\cmidrule(lr){6-7}\cmidrule(lr){8-9}
   & Gemini & Qwen & Gemini & Qwen & Gemini & Qwen & Gemini & Qwen \\
  \midrule
  II & 0.8164 & 0.8137 & 0.9613 & 0.9646 & 0.0930 & 0.0928 & 0.2272 & 0.2278 \\
  Silhouette & 0.8292 & 0.8238 & 0.9656 & 0.9597 & 0.0915 & 0.0953 & 0.2192 & 0.2226 \\
  Chromaticity & 0.8140 & 0.8136 & 0.9604 & 0.9632 & 0.0978 & 0.0971 & 0.2251 & 0.2262 \\
  Materiality & 0.8255 & 0.8230 & 0.9632 & 0.9629 & 0.0890 & 0.0899 & 0.2193 & 0.2204 \\
  Craftsmanship & 0.8063 & 0.8009 & 0.9611 & 0.9621 & 0.0892 & 0.0898 & 0.2146 & 0.2167 \\
  Wearability & 0.8291 & 0.8317 & 0.9639 & 0.9648 & 0.0906 & 0.0878 & 0.2194 & 0.2178 \\
  \bottomrule
  \end{tabular}}
  \end{table}
  
  \begin{table}[htbp]
  \centering
  \caption{Outfit-level consistency validation results }
  \label{tab:validation_comparison}
  \scalebox{0.7}{
  \begin{tabular}{lcccccccc}
  \toprule
  \multirow{2}{*}{Dimensions} & \multicolumn{2}{c}{Pearson $r$} & \multicolumn{2}{c}{Spearman $\rho$} & \multicolumn{2}{c}{MAE} & \multicolumn{2}{c}{RMSE} \\
  \cmidrule(lr){2-3}\cmidrule(lr){4-5}\cmidrule(lr){6-7}\cmidrule(lr){8-9}
   & Gemini & Qwen & Gemini & Qwen & Gemini & Qwen & Gemini & Qwen \\
  \midrule
  OI & 0.9700 & 0.9700 & 0.9652 & 0.9699 & 0.0861 & 0.0863 & 0.1095 & 0.1094 \\
  Silhouette & 0.9608 & 0.9615 & 0.9651 & 0.9663 & 0.0796 & 0.0799 & 0.1116 & 0.1109 \\
  Chromaticity & 0.9344 & 0.9323 & 0.9628 & 0.9620 & 0.0979 & 0.1007 & 0.1497 & 0.1279 \\
  Materiality & 0.9513 & 0.9462 & 0.9720 & 0.9621 & 0.0848 & 0.0892 & 0.1147 & 0.1204 \\
  Craftsmanship & 0.9630 & 0.9604 & 0.9666 & 0.9680 & 0.0694 & 0.0724 & 0.0908 & 0.0939 \\
  Wearability & 0.9572 & 0.9565 & 0.9667 & 0.9644 & 0.1009 & 0.1011 & 0.1254 & 0.1262 \\
  Stylistic Synergy & 0.9453 & 0.9466 & 0.9719 & 0.9668 & 0.0892 & 0.0924 & 0.1260 & 0.1244 \\
  Visual Harmony & 0.9555 & 0.9546 & 0.9685 & 0.9645 & 0.1189 & 0.1205 & 0.1559 & 0.1575 \\
  \bottomrule
  \end{tabular}}
  \end{table}

The validation results establish the credibility of LLM-based aesthetic assessment for fashion datasets. The strong correlation coefficients (Pearson $r>0.79$, Spearman $\rho>0.78$) and low error rates (MAE<0.7) across both general aesthetics and fashion-specific domains demonstrate that LLMs can reliably capture human aesthetic preferences. This enables us to employ Qwen-VL-Max for large-scale annotation of the POG dataset while maintaining annotation quality comparable to expert human annotators.
In addition to the two models mentioned above, we also conduct preliminary performance evaluations on several other multimodal large models, such as Tencent Hunyuan and Wenxin Yiyan. The results show that these models are not suitable for scoring this experiment, so they are not included in the experimental report.

\subsection{Large-scale Dataset Scoring Based on VLM}

After verifying the reliability of the scoring, we finally selected Qwen-VL-Max as the final scoring model. To ensure scoring accuracy and consistency, we carefully designed structured prompts. For input, we simultaneously input images of all items in the outfit and provide product titles from the dataset as image descriptions to the model, helping the model better understand clothing attributes and characteristics. For scoring requirements, we require the model to first independently score each item in the outfit according to 6 dimensions, then score the complete outfit according to 8 dimensions. The scoring range for each dimension is 1.0-10.0 points, retaining one decimal place, with higher scores indicating better aesthetic quality in that dimension.

\begin{figure*}[h]
\centering
\includegraphics[width=0.9\textwidth]{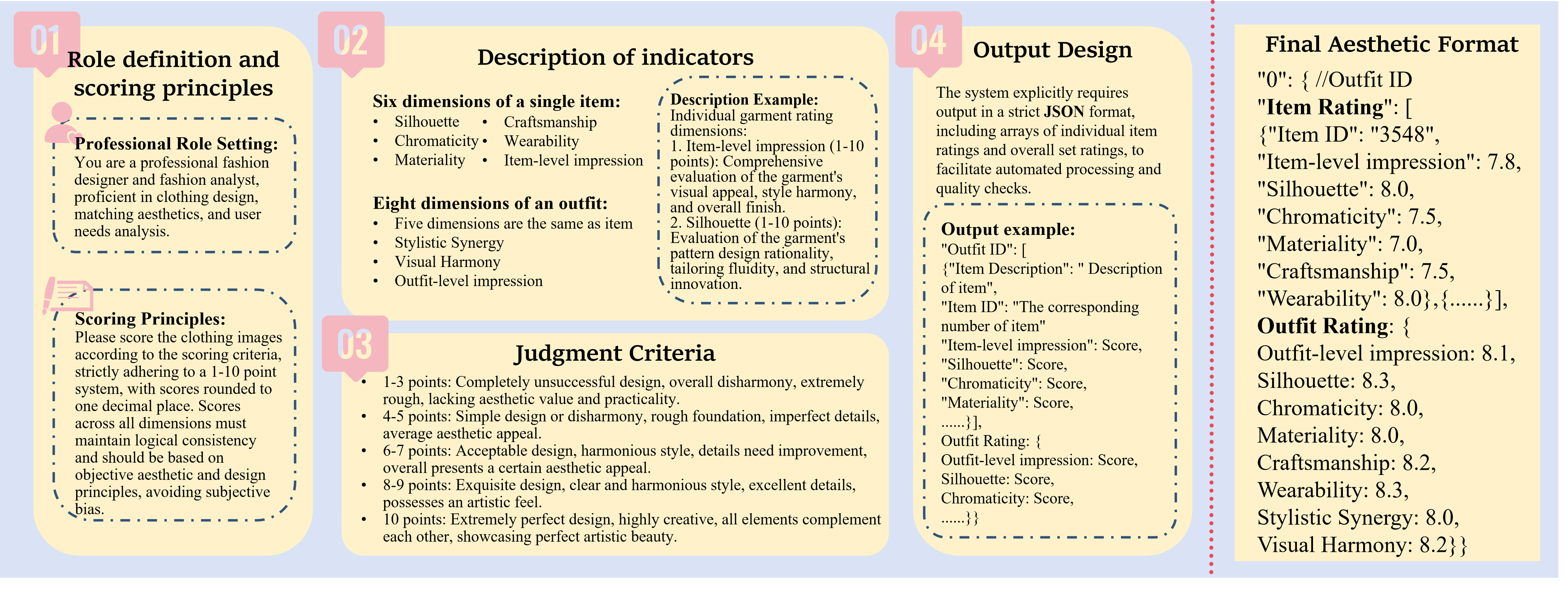}
\caption{Prompt Design and Final Aesthetic Format}
\label{fig:Prompt design}
\end{figure*}

To facilitate subsequent data processing and analysis, we require the model to output scoring results in structured JSON format. The output format contains two main parts: the \texttt{item\_scores} part records scores for each item in the outfit across 6 dimensions, and the \texttt{outfit\_scores} part records scores for the entire outfit across 8 dimensions. This structured output format ensures both the completeness of scoring data and facilitates automated processing and quality checking, ensuring the efficiency and accuracy of large-scale annotation work. The prompt design and the format of the aesthetic quality score in the final datasetis shown in Figure \ref{fig:Prompt design}.

\subsection{Dataset Statistical Analysis}

This section conducts systematic statistical analysis of the basic characteristics of the AesRec dataset. We conduct quantitative analysis of the dataset from two dimensions: data scale and aesthetic quality to validate its applicability in supporting aesthetics-aligned recommendation research.

\subsubsection{Data Scale and Interaction Density}

Table~\ref{tab:aesrec_statistics} shows the basic statistical information of the AesRec dataset. The dataset contains 53,843 users, 27,694 outfits, and 42,526 items, covering 17 semantic categories. At the interaction level, the dataset records 1,679,708 user-outfit interactions and 2,290,645 user-item interactions, with interaction densities reaching 0.11\% and 0.10\% respectively. There are 106,860 outfit-item affiliation relationships, reflecting the internal structural information of outfits. Furthermore, the size of the sets shows a clear concentration trend: 4-piece sets account for the highest proportion, reaching 39.73\%; 3-piece sets follow closely behind, accounting for 37.11\%; the two combined account for 76.84\%, forming the main body of the dataset. 5-piece sets account for 23.10\%, while extremely large sets are relatively rare—2-piece sets account for only 0.06\%. This scale is comparable to mainstream bundle recommendation datasets and can provide sufficient training samples for deep learning models.

\begin{table}[htbp]
  \centering
  \caption{AesRec Dataset Statistics}
  \label{tab:aesrec_statistics}
  \scalebox{0.8}{
  \begin{tabular}{lc|lc} 
  \toprule
  \textbf{Metric} & \textbf{Statistic} & \textbf{Metric} & \textbf{Statistic} \\
  \midrule
  \# Users & 53,843 & \# User-Item Inters. & 2,290,645 \\
  \# Outfits & 27,694 & \# Outfit-Item Affil. & 106,860 \\
  \# Items & 42,526 & U-I Interaction Density & 0.10\% \\
  \# User-Outfit Inters. & 1,679,708 & U-O Interaction Density & 0.11\% \\
  \midrule
  \multicolumn{4}{c}{\textit{Outfit Composition Details}} \\
  \midrule
  Avg. Items per Outfit & 3.86 & 3-piece Set (\%) & 37.11\% \\
  Outfit Size Range & 2--5 & 4-piece Set (\%) & 39.73\% \\
  2-piece Set (\%) & 0.06\% & 5-piece Set (\%) & 23.10\% \\
  \bottomrule
  \end{tabular}}
  \end{table}

\subsubsection{Aesthetic Quality Analysis}

\begin{figure}[htbp]
  \centering
  \includegraphics[width=0.45\textwidth]{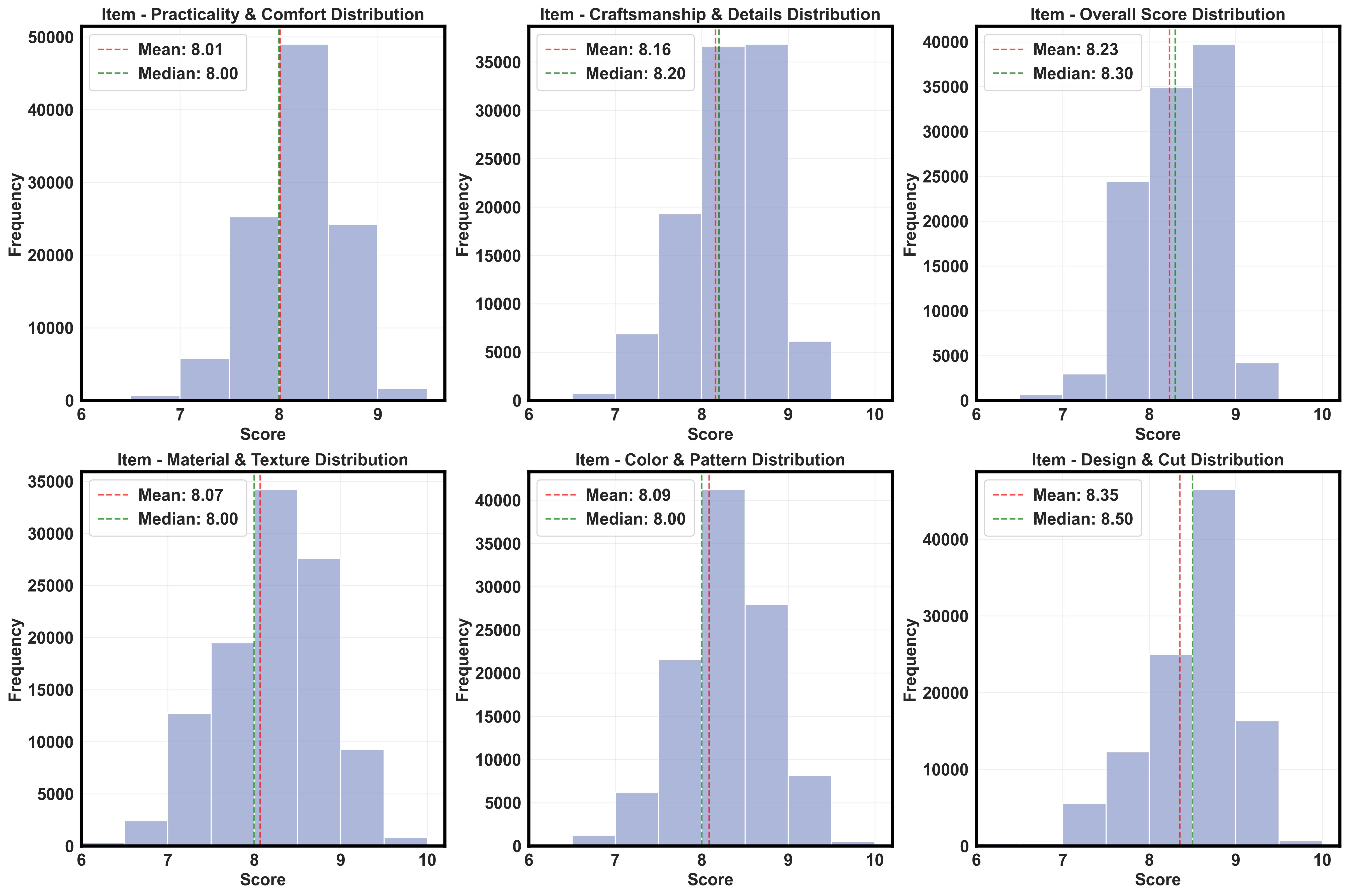}
  \caption{Comparison of aesthetic score distributions}
  \label{fig:aesthetic-item}
\end{figure}

\begin{figure}[htbp]
  \centering
  \includegraphics[width=0.45\textwidth]{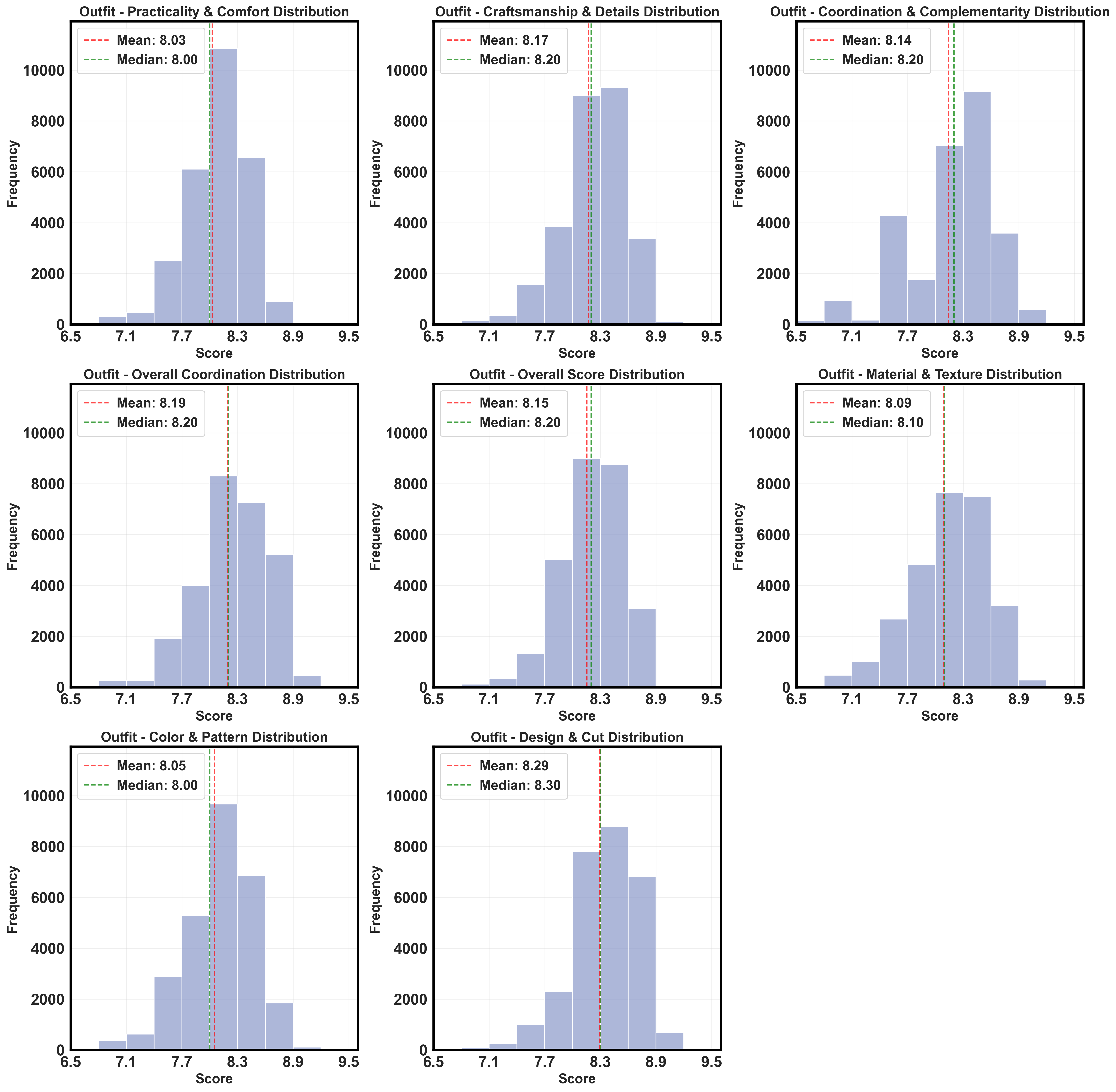}
  \caption{Comparison of aesthetic score distributions}
  \label{fig:aesthetic_outfit}
\end{figure}

Figure~\ref{fig:aesthetic-item} and \ref{fig:aesthetic_outfit} compares the distribution of outfit scores and item scores in histogram form. Across both individual items and complete outfits, scores across all aesthetic dimensions exhibit a distinct positive skew, with means and medians consistently ranging between $8.0$ and $8.5$. This indicates that the dataset overall maintains a high standard of aesthetic quality. The distribution curves for various dimensions—such as chromaticity, materiality, and craftsmanship—are highly similar and concentrated. This suggests that the large language model maintains a stable judgmental scale during multidimensional aesthetic assessment, systematically capturing both the global and local aesthetic nuances of clothing ensembles.

\begin{table}[htbp]
\centering
\caption{Aesthetic quality statistics}
\label{tab:aesthetic_stats}
\begin{tabular}{@{}lrrrrr@{}}
\toprule
Scoring Object & Mean & Median & Std Dev & Min & Max \\
\midrule
Outfit-level Impression  & 8.15 & 8.20 & 0.33 & 6.5 & 9.0 \\
Item-level Impression & 8.22 & 8.20 & 0.28 & 7.0 & 9.2 \\
\bottomrule
\end{tabular}
\end{table}

Table~\ref{tab:aesthetic_stats} provides detailed statistics in II and OI dimension. Quantile analysis shows that  more than three-quarters of outfits score above 8.0. This high-score concentrated distribution pattern reflects the overall high aesthetic quality level of the dataset. Item scores and outfit scores demonstrate good consistency. The mean difference between the two is only 0.07 points, standard deviations are similar, and medians are completely consistent. This consistency has important significance: on one hand, it validates the stability of the scoring system—whether evaluating individual items or complete outfits, the model can maintain consistent evaluation standards; on the other hand, it indicates that the overall aesthetic quality of outfits is indeed built on the foundation of item quality, conforming to the intuitive understanding that ``good outfits are composed of good items.

\section{Task Definition and Baseline Methods}

Although AesRec can support various recommendation tasks, this paper focuses on aesthetics-aligned recommendation, where quantitative aesthetic evaluation plays a crucial role in guiding users toward higher-quality outfit choices, while existing datasets lack such aesthetic annotations. 

\textbf{Formal Definition:}

\textbf{Input:} The input consists of two parts: (1) \textbf{Personalization components}: User set $U$, outfit set $O$, item set $I$, category set $C$, user-outfit interaction $R_{UO}$, user-item interaction $R_{UI}$, outfit-item composition relationship $R_{OI}$, item-category mapping $R_{IC}$; (2) \textbf{Aesthetic components}: Item-level aesthetic scores $A_I$ (6 dimensions: silhouette, chromaticity, materiality, craftsmanship, wearability, item-level impression), outfit-level aesthetic scores $A_O$ (8 dimensions: the 5 shared dimensions plus stylistic synergy, visual harmony, outfit-level impression).

\textbf{Output:} For each user $u \in U$, learn a personalized and aesthetics-aligned scoring function $f_u: O \rightarrow \mathbb{R}$ that integrates both personalization signals from interaction data and aesthetic scores $A_I$ and $A_O$ for ranking candidate outfits.

\section{Personalized Aesthetics Experiments}

\subsection{Experimental Setup}

All-unrated-item\cite{steck2013evaluation} as a widely used
evaluation protocol is adopted in this paper: for each user, all candidate outfits not interacted with in the training set are retained as the recommendation pool.
The experimental task is defined as Top-K outfit recommendation: generating a ranked recommendation list containing K outfits from candidate outfits for users.
The dataset is divided into training set, validation set, and test set at a ratio of 8:1:1.

\subsection{Evaluation Metrics}

\textbf{1) Personalized metrics:}Widely used Recall@K and NDCG@K are adopted to evaluate the personalization performance of Top-K recommendation~\cite{herlocker2004evaluating,shani2011evaluating,jarvelin2002cumulated}.
Recall measures the recall proportion of test outfits in the Top-K recommendation list, and NDCG considers the importance of recommendation positions by assigning higher weights to higher-ranked hits.
This paper sets K=10, 20, 40, 80 for evaluation.

\textbf{2) Aesthetic metrics:} The average aesthetic score of the Top-K recommendation list (denoted as $Score_{Aes}@K$) is adopted to evaluate the absolute aesthetic level of recommendation results~\cite{talebi2018nima,deng2017image}.
The calculation formula is as follows:

\begin{equation}
\label{eq:avg_aesthetic_score}
Score_{Aes}@K = \frac{\sum\{A_m(m_i) \mid m_i \in RL\}}{|RL|}
\end{equation}

where $RL$ represents the recommendation list, and $A_m(m_i)$ represents the aesthetic score of outfit $m_i$. 

In addition, we introduce a relative aesthetic improvement indicator $\Delta Score_{Aes}@K$, defined as the difference between the aesthetic score of the recommendation list and the aesthetic score of user historical interacted outfits: 

\begin{equation}
\label{eq:relative_aesthetic_improvement}
\Delta Score_{Aes}@K = Score_{Aes}@K - Score_{historic}
\end{equation}

where $Score_{historic}$ is the average aesthetic score of historical data.
A positive $\Delta Score_{Aes}@K$ indicates that the aesthetic level of the recommendation list is higher than user historical preferences, while a negative value indicates the opposite.

\begin{table*}[htbp]
  \centering
  \caption{Personalized Recommendation Performance Comparison}
  \label{tab:personalization_performance}
  \small
  \begin{tabular}{lcccccccc}
  \toprule
  \multirow{2}{*}{\textbf{Model}} & \multicolumn{2}{c}{$K=10$} & \multicolumn{2}{c}{$K=20$} & \multicolumn{2}{c}{$K=40$} & \multicolumn{2}{c}{$K=80$} \\
  \cmidrule(lr){2-3}\cmidrule(lr){4-5}\cmidrule(lr){6-7}\cmidrule(lr){8-9}
   & Recall$\uparrow$ & NDCG$\uparrow$ & Recall$\uparrow$ & NDCG$\uparrow$ & Recall$\uparrow$ & NDCG$\uparrow$ & Recall$\uparrow$ & NDCG$\uparrow$ \\
  \midrule
  BPR-MF & 0.0421 & 0.0318 & 0.0689 & 0.0417 & 0.1081 & 0.0538 & 0.1658 & 0.0686 \\
  HyperMBR & 0.0611 & 0.0473 & 0.0959 & 0.0604 & 0.1453 & 0.0755 & 0.2108 & 0.0923 \\
  MIDGN & 0.0422 & 0.0325 & 0.0680 & 0.0421 & 0.1058 & 0.0537 & 0.1592 & 0.0674 \\
  MultiCBR & 0.1055 & 0.0884 & 0.1487 & 0.1045 & 0.2016 & 0.1207 & 0.2650 & 0.1369 \\
  CrossCBR & 0.0785 & 0.0636 & 0.1174 & 0.0781 & 0.1672 & 0.0934 & 0.2310 & 0.1098 \\
  BGCN & 0.0505 & 0.0386 & 0.0806 & 0.0499 & 0.1246 & 0.0634 & 0.1875 & 0.0795 \\
  LLM-qwen & 0.1791 & 0.1344 & 0.2043 & 0.1425 & 0.2618 & 0.1579 & 0.3042 & 0.1675 \\
  AesRec & 0.0659 & 0.0532 & 0.0990 & 0.0656 & 0.1446 & 0.0796 & 0.2040 & 0.0948 \\
  \bottomrule
  \end{tabular}
  \end{table*}
  
  \begin{table*}[htbp]
  \centering
  \caption{Performance Comparison: Aesthetic Scoring and Ranking Exposure}
  \label{tab:performance_comparison}
  \small
  \begin{tabular}{lcccccccc}
  \toprule
  \multirow{2}{*}{\textbf{Model}} & \multicolumn{2}{c}{$K=10$} & \multicolumn{2}{c}{$K=20$} & \multicolumn{2}{c}{$K=40$} & \multicolumn{2}{c}{$K=80$} \\
  \cmidrule(lr){2-3}\cmidrule(lr){4-5}\cmidrule(lr){6-7}\cmidrule(lr){8-9}
   & $Score_{Aes}\uparrow$ & ExpoGap$\uparrow$ & $Score_{Aes}\uparrow$ & ExpoGap$\uparrow$ & $Score_{Aes}\uparrow$ & ExpoGap$\uparrow$ & $Score_{Aes}\uparrow$ & ExpoGap$\uparrow$ \\
  \midrule
  BPR-MF & 8.1121(-0.0370) & -0.1190 & 8.1197(-0.0294) & -0.1069 & 8.1278(-0.0214) & -0.0958 & 8.1353(-0.0139) & -0.0860 \\
  HyperMBR & 8.1307(-0.0193) & -0.0295 & 8.1311(-0.0178) & -0.0288 & 8.1328(-0.0136) & -0.0277 & 8.1392(-0.0097) & -0.0268 \\
  MIDGN & 8.1311(-0.0181) & -0.1320 & 8.1334(-0.0157) & -0.1170 & 8.1377(-0.0115) & -0.1036 & 8.1420(-0.0071) & -0.0926 \\
  MultiCBR & 8.1440(-0.0051) & -0.0412 & 8.1476(-0.0016) & -0.0350 & 8.1500(+0.0008) & -0.0307 & 8.1505(+0.0013) & -0.0283 \\
  CrossCBR & 8.1373(-0.0119) & -0.0537 & 8.1412(-0.0079) & -0.0488 & 8.1437(-0.0055) & -0.0453 & 8.1453(-0.0038) & -0.0428 \\
  BGCN & 8.1439(-0.0053) & -0.1249 & 8.1455(-0.0036) & -0.1316 & 8.1478(-0.0013) & -0.1346 & 8.1493(+0.0002) & -0.1389 \\
  LLM-qwen & 8.1227(-0.0137) & -0.0952 & 8.1238(-0.0126) & -0.0890 & 8.1247(-0.0113) & -0.0889 & 8.1273(-0.0086) & -0.0913 \\
  AesRec & 8.3561(+0.2070) & +0.6486 & 8.3405(+0.1913) & +0.6159 & 8.3229(+0.1738) & +0.5829 & 8.3031(+0.1539) & +0.5492 \\
  \bottomrule
  \end{tabular}
  \end{table*}

\textbf{2) Ranking Exposure Metrics:} Considering that highly ranked outfits are more likely to receive user clicks, Ranking Exposure is adopted to evaluate the ranking performance of outfits with different aesthetic levels~\cite{diaz2020evaluating,singh2018fairness,zehlike2017fa}.
Outfits are ranked by aesthetic level, with the first half classified into the group of relatively higher aesthetic level $G_h$, and the second half classified into the group of relatively lower aesthetic level $G_{lh}$.
The ranking exposure of outfit group $G_h$ under ranking $\pi$ is defined as:

\begin{equation}
\label{eq:ranking_exposure}
Expo(G_h|\pi) = \sum_{m \in G_h} Utility(m|\pi)
\end{equation}

where $Utility(\cdot)$ represents a utility function, typically a monotonically decreasing function with respect to outfit rank.
This paper sets $Utility(m|\pi) = [rank(m|\pi)]^{-1}$.
The ranking exposure gap between groups $G_h$ and $G_{lh}$ under ranking $\pi$ is defined as:

\begin{equation}
\label{eq:exposure_gap}
ExpoGap(G_h, G_{lh}|\pi) = \frac{Expo(G_h|\pi) - Expo(G_{lh}|\pi)}{Expo(G_h|\pi) + Expo(G_{lh}|\pi)}
\end{equation}

We abbreviate $ExpoGap(G_h, G_{lh}|\pi)$ as $ExpoGap$.
Since both groups have the same size, $ExpoGap = 0$ indicates that there is no difference in ranking exposure between the two groups.
$ExpoGap < 0$ ($ExpoGap > 0$) indicates that outfits with relatively lower (higher) aesthetic level have higher ranking exposure, with the numerical value indicating the degree of gap.

\subsection{Baseline and Hyperparameter Settings}

We select several state-of-the-art models that have demonstrated superior performance in the fields of clothing recommendation and outfit recommendation as our evaluation baselines. 
These models are mainly divided into three categories: (1) \textbf{Independent interaction learning methods}: BGCN~\cite{he2020bgcn} learns user preferences separately at item and outfit levels, while BPR-MF~\cite{rendle2009bpr} only conducts interaction learning at the outfit level using matrix factorization; 
(2) \textbf{Collaborative interaction learning methods}: MultiCBR~\cite{ma2024multicbr} employs multi-task learning, HyperMBR~\cite{ke2023hyperbolic} achieves mutual learning in hyperbolic space, CrossCBR~\cite{ma2022crosscbr} utilizes contrastive learning, and MIDGN~\cite{zhao2022midgn} adopts an intermediate layer graph network; 
(3) \textbf{Large language model recommendation methods}: LLM-based methods leverage pre-trained language models for recommendation. All baseline models do not use aesthetic scoring information to ensure fair comparison. All baseline models follow the parameters in the original paper.

\subsection{Baseline Performance}

We first analyze the performance of baseline models that do not incorporate aesthetic guidance mechanisms. These models focus solely on learning user preferences from historical interaction data, without explicit aesthetic optimization objectives.

Our analysis reveals three key observations: (1) \textbf{Personalization Performance}: As shown in Table~\ref{tab:personalization_performance}, baseline models demonstrate varying but generally reasonable personalization capabilities, with MultiCBR achieving Recall@20 and NDCG@20 of 0.1487 and 0.1045 respectively. The superior personalization performance of LLM-qwen in clothing outfit recommendation may be attributed to the candidate set construction approach.

(2) \textbf{Aesthetic Quality Deficiency}: Table~\ref{tab:performance_comparison} reveals that all baseline models exhibit negative or near-zero $\Delta$AestheticScore@K values across different K settings, indicating systematic degradation in aesthetic quality. Even advanced collaborative learning models like MultiCBR and BGCN achieve only marginal improvements (e.g., +0.0013 and +0.0002 at K=80), while simpler models such as BPR-MF show significant aesthetic degradation ($\Delta$Score = -0.0370 at K=10). This demonstrates that sophisticated collaborative learning frameworks cannot effectively address aesthetic quality without dedicated aesthetic optimization objectives. 

(3) \textbf{Ranking Bias}: Table~\ref{tab:performance_comparison} shows that all baseline models consistently exhibit negative ExpoGap values, revealing a systematic bias toward low-aesthetic outfits. The bias magnitude ranges from -0.0288 (HyperMBR) to -0.1320 (MIDGN at K=10), indicating that even the best-performing baseline models fail to ensure fair exposure for high-aesthetic content. 

These findings collectively highlight the critical limitation of existing recommendation approaches in maintaining aesthetic quality and underscore the necessity of incorporating explicit aesthetic guidance mechanisms.

\section{Improve Aesthetic of Recommendation}

To address the aesthetic quality deficiency identified in baseline models, we propose AesRec, an aesthetic-enhanced recommendation method built upon CrossCBR. AesRec explicitly incorporates aesthetic scoring information into the recommendation framework through a joint loss function that simultaneously optimizes for both personalization and aesthetic quality. Following the joint loss formulation approach~\cite{li2024mealrec+}, we combine the personalized recommendation loss from CrossCBR with an aesthetic preference ranking loss, enabling the model to learn user preferences while promoting higher aesthetic quality in recommendations.

The joint loss function $\mathcal{L}$ is formulated as follows:
\begin{equation}
\mathcal{L} = \mathcal{L}_{\text{rec}} + \lambda \mathcal{L}_{\text{aesthetic}}
\end{equation}
where $\mathcal{L}_{\text{rec}}$ denotes the personalized recommendation loss inherited from CrossCBR, which captures user preferences based on historical interactions, and $\mathcal{L}_{\text{aesthetic}}$ represents the aesthetic preference ranking loss that encourages the model to rank outfits with higher aesthetic scores higher in the recommendation list. The hyperparameter $\lambda$ controls the trade-off between personalization and aesthetic quality, allowing us to balance these two objectives effectively.

The aesthetic preference ranking loss $\mathcal{L}_{\text{aesthetic}}$ is designed to ensure that outfits with higher aesthetic scores receive higher ranking positions. Specifically, for a user $u$ and a pair of outfits $(o_i, o_j)$ where outfit $o_i$ has a higher aesthetic score than outfit $o_j$, the loss penalizes cases where the model's predicted preference score for $o_i$ is lower than that for $o_j$. This formulation aligns the recommendation ranking with aesthetic quality, ensuring that users are exposed to aesthetically superior content while maintaining personalized relevance.

 Table~\ref{tab:personalization_performance} and Table~\ref{tab:performance_comparison} show the comprehensive comparison results across personalized recommendation, aesthetic scoring, and ranking exposure metrics.

\textbf{AesRec maintains competitive personalized recommendation performance while achieving substantial aesthetic improvements.} As shown in Table~\ref{tab:personalization_performance}, AesRec achieves Recall@20 and NDCG@20 of 0.0990 and 0.0656 respectively, slightly lower than MultiCBR (0.1487, 0.1045) but significantly better than BPR-MF, MIDGN, and BGCN. Compared to CrossCBR, AesRec's Recall@20 decreases from 0.1174 to 0.0990, a limited trade-off justified by significant aesthetic improvements, validating the joint loss function design. Furthermore, AesRec is evaluated without hyperparameter tuning, adopting the same hyperparameters as CrossCBR, indicating potential for further improvement in personalization performance. 

\textbf{AesRec achieves substantial aesthetic improvements across all K settings.} Table~\ref{tab:performance_comparison} shows that AesRec achieves consistently positive $\Delta$AestheticScore@K values (+0.2070 to +0.1539), with $\Delta$Score of +0.1913 at K=20, representing a 2.4\% improvement over CrossCBR. Notably, AesRec's improvement magnitude (+0.19) is approximately 24 times larger than the best baseline (MultiCBR's +0.0013), demonstrating the advantage of explicit aesthetic guidance.

\textbf{AesRec achieves dramatic ranking exposure reversal, with improvement magnitude exceeding aesthetic score gains.} Table~\ref{tab:performance_comparison} shows AesRec achieves positive ExpoGap values (+0.6486 to +0.5492), with ExpoGap of +0.6159 at K=20, representing a 0.6647 improvement over CrossCBR's -0.0488. At K=20, AesRec assigns 2.9069 exposure to high-aesthetic outfits versus 0.6909 to low-aesthetic ones (4.21:1 ratio), while CrossCBR shows the opposite pattern (1.7110 vs 1.8867). The ExpoGap improvement (0.6647) is 3.5 times larger than aesthetic score improvement (0.19), indicating that aesthetic preference ranking loss effectively optimizes ranking positions, ensuring high-aesthetic outfits occupy favorable exposure positions.

\section{Conclusion and Future Work}

This paper constructs the AesRec benchmark dataset with systematic quantitative aesthetic annotations to address the critical gap in aesthetics-aligned clothing recommendation. Grounded in internationally recognized apparel quality standards, AesRec provides multidimensional aesthetic scoring across eight dimensions for outfits and six dimensions for individual items. Using advanced vision-language models for large-scale aesthetic scoring with rigorous human-model consistency validation, the dataset contains 27,694 outfits, 42,526 items, and 1.68 million user-outfit interactions. Experimental results demonstrate that incorporating quantitative aesthetic information significantly improves recommendation aesthetic quality while maintaining competitive personalized performance. The proposed AesRec method, integrating aesthetic scoring through a joint loss function, achieves substantial improvements in aesthetic metrics and ranking exposure fairness. Future directions include incorporating temporal dynamics, exploring personalized aesthetic preferences, investigating cross-cultural variations, and developing more efficient annotation methods. The AesRec dataset provides a foundation for advancing aesthetics-aligned recommendation research.

\newpage
\bibliographystyle{unsrtnat}
\bibliography{references}

\end{document}